# FAST LASER COOLING OF LONG LIVED ION BEAMS

E.G. Bessonov, A.L. Osipov, Lebedev Phys. Inst. RAS, Moscow, Russia
A.A. Mikhailichenko, Cornell University, CLASSE, Ithaca, NY 14853, U.S.A.

*Abstract*

Some peculiarities of fast laser cooling of long-lived ion beams in storage rings are discussed. Selective interaction of ions and broadband laser beam with sharp frequency and geometric edges is used while laser and ion beams are partially overlapped. The rates of change of the ion beam density in different regions of the phase space and at different moments of time in this scheme of cooling differ. That is why the generalized Robinson theorem valid for the infinitesimal phase space regions of non exponential cooling in turn is used to interpret the results.

## INTRODUCTION

Laser ion cooling (LIC) is a well-known technique now [1]-[4]. Basic idea of this method is in irradiation of moving ion beam with a laser, having frequency which if transformed into the moving frame becomes close to the transition levels. As the ion is moving, the photon radiated after the ion jumps into the ground state will carry substantial momenta directed along the instant direction of motion, while transformed into the Lab frame.

Other well-known technique is so called Enhanced Optical Cooling (EOC). Its basic idea is in application the cooling procedure to the fraction of the beam, first- to the particles with the highest deviation from equilibrium and further on to the ones with lower deviations. To manipulate with amplified optical signal from pick-up undulator we suggested an electro-optical deflection device or even mechanical system for slower processes [5].

In this report we considered the benefits of implementation of EOC technique to the LIC. More exactly we suggest irradiating just fraction of the moving ion beam at a time and further on expanding illumination area to ~ half of the cross section of ion beam. Below we describe this idea in detail.

The trajectory of a particle in external fields is described by the equation $d\vec{p}/dt = \vec{F}_{ext} + \vec{F}_{fr}$, where $\vec{p} = m\gamma\vec{v}$ is the particle momentum, $\vec{F}_{ext} = e(\vec{E}_{ext} + (1/c)[\vec{v}\vec{H}_{ext}])$ is the external force, $\vec{F}_{fr} = F_{fr}(\vec{r},\vec{v})\vec{n}_v$ is the frictional force, $m$ is the particle rest mass (the mass depends on its state: excited, non excited), $\gamma = \varepsilon/mc^2$ is its relativistic factor, $\vec{n}_v = \vec{v}/v$ is the unit vector in the direction of the particle velocity, $\varepsilon$ is the total energy of the particle, $\vec{r}$ is the radius vector, $\vec{v}$ is the vector of particle velocity, $v = |\vec{v}|$, $F_{fr} = |\vec{F}_{fr}|$, $\vec{E}_{ext}$ and $\vec{H}_{ext}$ are the external electric and magnetic fields. In this case the six dimensional infinitesimal volume of an ensemble of particles in a 6D phase space region gathered round some selected particle is decreased by the law $\rho = \rho_0 \exp[-\int \alpha_{6D}(\vec{r},\varepsilon,t) \, dt]$ with the instantaneous decrement determining the rate of cooling

$$\alpha_{6D}(\vec{r},\varepsilon,t) = (1+\frac{1}{\beta^2})\frac{P_{fr}(\vec{r},\varepsilon,t)}{\varepsilon} + \frac{\partial P_{fr}(\vec{r},\varepsilon,t)}{\partial \varepsilon}, \quad (1)$$

where $P_{fr} = \vec{F}_{fr}\vec{v} = F_{fr}v$ is the power of the particle energy loss [1].

In synchrotrons or storage rings the energy losses of ions can be recovered by their RF systems (cooling in a bucket). The 6D increment of such beam in this case is the sum of radial, vertical and longitudinal increments: $\alpha_{6D} = 2\alpha_x + 2\alpha_z + 2\alpha_s$. In the relativistic case:

$$\alpha_x = \frac{1}{2}\left[\frac{\overline{P_s}}{\varepsilon_s} + \frac{\partial \overline{P}}{\partial \varepsilon}|_s - \frac{d\overline{P}}{d\varepsilon}|_s\right], \quad \alpha_z = \frac{1}{2}\frac{\overline{P_s}}{\varepsilon_s}, \quad \alpha_l = \frac{1}{2}\frac{d\overline{P}}{d\varepsilon}|_s, \quad (2)$$

where the over lined values are the values averaged over many periods of the particle revolution in the ring, $\varepsilon_s$ is the equilibrium ion energy in the ring [1]. The first and second terms in (2) are related to the damping of the radial and vertical betatron oscillations in the transverse plane and the third one $\overline{\alpha_l}$ is related to the damping of the phase oscillations in the longitudinal plane. It is supposed that partial increments (2) are valid inside all area of the corresponding planes. The six dimensional (6D) damping time of the ion beam in this case is:

$$\tau_{6D} = 1/\overline{\alpha_{6D}}. \quad (3)$$

If relativistic particle beam emits synchrotron and undulator radiation then the power $\overline{P}_{Fr}(\varepsilon) = k_{fr}\varepsilon^2$, the radiative damping decrement $\overline{\alpha_{6D}} = 4\overline{P}_{fr}(\varepsilon)/\varepsilon|_{\varepsilon=\varepsilon_s}$, where $k_{fr}$ is a constant. The decrement for the ionization cooling based on the energy loss of particles in a matter targets has more complicated nonlinear form [2]. Cooling based on such energy losses has the partial derivatives $\partial \overline{P}_{fr}(\varepsilon)/\partial \varepsilon \sim \overline{P}_{fr}(\varepsilon)/\varepsilon$ and small decrements $\alpha_{6D}$.

In the case of radiative cooling of ion beams in storage rings by broadband lasers [3], [4] the target is the contra propagated laser beam, where the backward Rayleigh scattering of laser photons takes place. The spectral distribution of the intensity of the laser beam $\overline{I}_{\omega,l}$ determines the partial derivative $\partial \overline{P}_{fr}(\varepsilon)/\partial \varepsilon$ and hence the rates of the ion beam cooling. For ordinary radiative ion beam cooling [3] the spectral distribution of the laser beam intensity is uniform in the frequency band $\Delta\omega_l = \omega_{l,max} - \omega_{l,min}$. It corresponds to the resonance conditions for the excitation of the ion beam in the limits of its energy spread $\Delta\varepsilon_b$. In this case the derivative

$\partial \overline{P}_{fr}(\varepsilon)/\partial\varepsilon \sim \overline{P}_{fr}(\varepsilon)/\varepsilon$ is small. For stimulated radiative cooling of ion beam the spectral distribution of the laser beam intensity is increased from zero to maximum value in the same frequency band and the power loss has linear increase from zero to maximum value in the energy band $\Delta\varepsilon_b$. In this case $\partial \overline{P}_{fr}(\varepsilon)/\partial\varepsilon \sim \overline{P}_{fr}(\varepsilon)/\Delta\varepsilon_b$ [4]. It means that the rate of the stimulated radiative ion cooling is $\varepsilon_s/\Delta\varepsilon_b \sim 10^2 - 10^3$ times higher than ordinary one.

## FAST LASER COOLING OF LONG LIVED ION BEAMS

Below the peculiarity of fast laser cooling of relativistic long lived ion beams based on their selective interaction with broadband, sharp frequency and geometric laser beam edges is discussed. The "cooling" goes on both in the longitudinal and transverse directions simultaneously.

*Cooling without a RF bucket*

First for simplicity, we suppose that the RF system of the ring is switched off, and the spectral distribution of the laser beam intensity is uniform in a frequency band $\Delta\omega_l = \omega_{l,\max} - \omega_{l,\min}$. The power of the ion energy loss is increased from zero at the ion energy $\varepsilon \le \varepsilon_c$ to maximum value $P_{fr,m}$ at the ion energy $\varepsilon > \varepsilon_c$ on a short energy interval $\delta\varepsilon \ll \Delta\varepsilon_b = \varepsilon_{\max} - \varepsilon_c$ and then stay constant up to the energy $\varepsilon_{\max}$ corresponding to the boundary frequency of the laser beam spectra $\omega_{l,\min} = \omega_{tr}/2\gamma_{\max}$ (step function), where $\omega_{tr}$ is the ion resonant transition frequency at rest, $\gamma_{\max} = \varepsilon_{\max}/Mc^2$, $\varepsilon_{\max}$ and $M$ are the maximal ion energy in the beam and its mass. The interaction region of the laser and ion beams is located in the straight section with zero dispersion function $\eta_{x,int} = 0$. Instantaneous orbits of ions in this section are overlapped and that is why the transverse dimensions of the beam are determined by betatron oscillations only.

Let the ion and laser beams are partially overlapped, laser beam is located at a negative deviation $\delta_{l,i}$ from the central ion orbit, ions with negative deviations from their instantaneous orbits only interact with the laser beam and become excited (see Fig. 1). Then the excited ions go out of the straight section of the ring and move in the region with positive dispersion function. Emission of photons by these ions in this region leads to the negative jumps of their instantaneous orbits in the direction of ions and hence to the decrease of their amplitudes of betatron oscillations (see Fig.2).

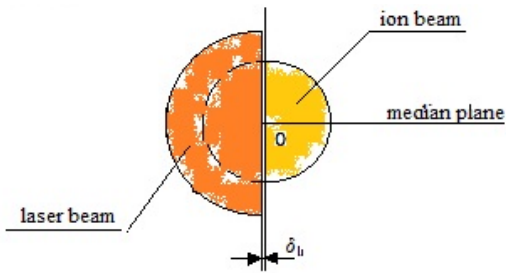

Fig.1. The interaction region of ion and laser beams.

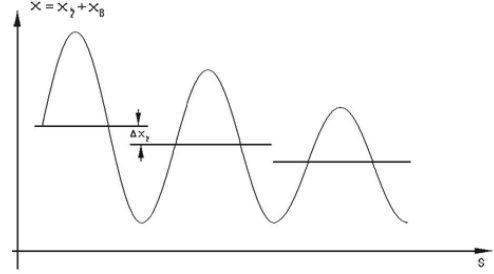

Fig.2. The evolution of the amplitude of an ion betatron oscillations in the radial plane.

Next turn another part of ions will have negative amplitudes of betatron oscillations, will be excited in the straight section and decrease their amplitudes in the region with positive dispersion function. Such a way every turn a part of ions of the beam decreases theirs amplitudes. If the ion decay time is lesser than its propagation time through the region with positive deviation from the instantaneous orbit ($\sim \lambda_b/4$, where $\lambda_b$ is the wavelength of the betatron oscillation) then the effective decrease of ion betatron oscillations will take place.

If average jump of ion instantaneous orbit in the region with positive dispersion function is $\overline{\delta} = \overline{P}_{fr}(\varepsilon)T\overline{\eta}_x/\varepsilon$ and maximal amplitude of betatron oscillations is $A_{\max}$ then the damping time of the ion beam in the radial (horizontal) plane will be equal to

$$\tau_x = 1/\overline{\alpha}_x \simeq 2T\frac{A_{\max}}{\overline{\delta}}, \qquad (4)$$

where $\overline{\eta}_x > 0$ is the average dispersion function in the region of the ion decay, $T$ is the revolution period. Linear (non-exponential) decrease of betatron oscillation amplitudes will occur from the value $A = A_{\max}$ to $A \simeq 0$. In this case the instantaneous orbits of ions having maximal amplitudes of betatron oscillations will pass the way $\Delta x_\eta \simeq A_{\max}$ and the instantaneous orbits of ions having near zero amplitudes of betatron oscillations will stay immovable. It means that radial dimensions of the ion beam in the straight section will be decreased to a small value $\delta_{l,i}$ (see Fig. 1) and the initial spread of the instantaneous orbits $\Delta x_{\eta 0}$ will be increased to the value $\Delta x_{\eta f} \simeq \Delta x_{\eta 0} + A_{\max}$.

After the decrease of the amplitudes of betatron oscillations of the ion beam in the radial plane and increase of its energy spread the beam can be cooled in the longitudinal plane. In this case the laser beam must be shifted on the right (geometric cutoff edges must be removed) to give to all ions an opportunity to interact with laser beam in the limits of its band $\Delta\omega_l = \omega_{l,\max} - \omega_{l,\min}$. The damping time of the beam in the longitudinal plane will be

$$\tau_l = 1/\overline{\alpha}_l = 2T\frac{\Delta x_{\eta,f}}{\overline{\delta}}. \qquad (5)$$

Fast laser cooling of ion beams in the longitudinal plane in this case takes place only at the boundary of the frequency spectra of the laser beam $\omega_{l,\max}$ in a short interval of the ion energy $\delta\varepsilon \ll \Delta\varepsilon_b$ and the corresponding interval of the laser frequency $\delta\omega \ll \Delta\omega_l$, where the partial derivative $\partial \overline{P}_{fr}(\varepsilon)/\partial\varepsilon$ is very high (step function is an idealization). The increase of betatron oscillations in this case will not occur as ions will be excited and emit radiation under conditions of both negative and positive deviations from their instantaneous orbit.

After cooling the ion beam can be accelerated to the necessary value by the RF system of the storage ring using phase displacement mechanism.

*Cooling in the RF bucket*

Fast laser cooling of long lived ion beams can work with RF system of the ring switched on also. In this case the emission of photons will lead to both fast damping of radial betatron oscillations and fast antidumping of phase oscillations. To remove the antidumping of ion phase oscillations the lower frequency edge of the laser beam ought to increase to the value a little lower $\omega_{l,s}$, where $\omega_{l,s}$ corresponds to the resonance frequency of interaction of electromagnetic wave with ion having the equilibrium energy $\varepsilon_s$. The spectral distribution of the laser beam can be a rising function of the difference ($\varepsilon - \varepsilon_s$). Stimulated radiative cooling system can be installed in another straight section of the ring with the nonzero dispersion function [4] as well to remove the antidumping in the longitudinal plane.

Note that: 1) Cooling effect may stay noticeable for the case of ion living time higher some periods of their betatron oscillations; 2) The laser beam dimensions can be decreased during the cooling process in accordance with the decrease of ion beam dimensions, 3) If the dispersion function in the straight section of the ring $\eta_x > 0$, maximal amplitudes of betatron oscillations $A_{x,m} \gg \delta_{l,i}$ then fast damping of amplitudes of ion betatron oscillations is possible for short lived ions. In this case the laser beam must cover the ion beam step-by-step when moving from negative position $x_\eta < 0$ to $x_\eta = -\delta_{l,i}$. First of all, it will interact with ions which have negative deviations from theirs instantaneous orbits $x_\eta < 0$ and decrease their amplitudes $A_x$ (emittance exchange). Fast cooling in longitudinal plane in the another straight section and the automatic repeat of this procedure must be used. 4) Instead of the considered laser beam with the sharp geometric edges the narrow focused laser beam moving in a rectangular area (from $x_\eta < 0$ to $x_\eta = -\delta_{l,i}$ and from $-z_m$ to $z_m$) in the form a series of vertically arranged lines can be used.

## CONCLUSION

1. The considered scheme of cooling works only for the ions with nonzero lifetime. 2. If the ion and laser beams were fully overlapped, then the selectivity would disappear and ions would have equal probability to be excited and decay at positive and negative deviations from their instantaneous orbits and will not change their amplitudes. 3. The generalized Robinson theorem valid for the infinitesimal phase space regions was used for the explanation of results in this in general case non exponential way of cooling. 4. Linear rate of the decrease of the betatron oscillation amplitudes in this scheme of cooling *is faster* than the exponential one. 5. The Laser beam edge must have a little negative radial deviation from the center of the ion beam ($\delta_{l,i} \ll$ ion beam diameter; see Fig 1). That is why the minimal residual amplitudes of betatron oscillations in this scheme are determined by the maximal jump of the instantaneous orbits $\overline{\delta}$ and alignment accuracy of laser and ion beams. 6. This scheme of cooling is similar to the enhanced optical cooling one [5]. It can present the additional possibilities to the existing schemes of cooling. 7. Fast damping of the radial betatron oscillations based on the step by step displacement of the laser beam in the straight section with the nonzero dispersion function $\eta_x > 0$ was suggested for short lived ions.

## REFERENCES

[1] E.G. Bessonov, The evolution of the phase space density of particle beams in external fields, Proceedings of COOL 2009, Lanzhou, China, p. 91-93, http://accelconf.web.cern.ch/AccelConf/COOL2009/papers/tua2mcio02.pdf; http://lanl.arxiv.org/abs/0808.2342.

[2] E.G.Bessonov, Methods of charged particle beam cooling, p. 606-618, in book "Charged and Neutral Particles Channeling Phenomena - Channeling 2008", Proceedings of the 51st Workshop of the INFN Eloisatron Project, S.B. Dabagov and L. Palumbo, Eds., World Scientific, 2010. (21 Oct. -1Nov. 2008, Erice, Italy).

[3] E.G.Bessonov, Kwang-Je Kim, Radiative cooling of ion beams in storage rings by broad band lasers, Phys. Rev. Lett., 1996, v.76, No 3, p.431-434.

[4] E.G.Bessonov, R.M.Feshchenko, Stimulated radiation cooling, Proc. XXI Russian Particle Accelerator

Conference, RuPAC-2008, Russia 2008, p. 91-92; http://cern.ch/AccelConf/r08/papers/MOCAU03.pdf .

[5] E.G. Bessonov, M.V. Gorbunkov, A.A. Mikhailichenko, Enhanced optical cooling system test in an electron storage ring. Phys. Rev. ST Accel. Beams **11**, 011302 (2008); http://lanl.arxiv.org/abs/0704.0870, http://prst-ab.aps.org/pdf/PRSTAB/v11/i1/e011302.